\documentclass[10pt,a4paper]{article}
\usepackage{amsmath,amssymb,graphicx,geometry,indentfirst,natbib,authblk,bm,hyperref,aas_macros}
\usepackage[font=small]{caption}
\usepackage[title]{appendix}
\hypersetup{colorlinks=true,linktoc=page,linkcolor=blue,citecolor=blue,urlcolor=blue,pdfstartview=FitH,bookmarksopen=true}
\setcitestyle{notesep={}}
\begin{document}
\title{\textbf{Convection Anisotropies of Cosmic Rays in Highly Magnetized Plasma}}
\author[1]{Yiran Zhang\footnote{\href{mailto:zhangyr@swjtu.edu.cn}{zhangyr@swjtu.edu.cn}}}
\author[1]{Siming Liu}
\affil[1]{School of Physical Science and Technology, Southwest Jiaotong University, Chengdu 611756, China}
\maketitle
\begin{abstract}
Recently, \cite{2024ApJ...964L...1Z} proposed a turbulent convection model for multiscale anisotropies of cosmic rays (CRs), with an assumption of isotropic diffusion such that the anisotropies are statistically isotropic. However, this assumption may be unrealistic for TeV CRs, whose observations have revealed the significance of the local interstellar background magnetic field. To meet the difficulty, the turbulent convection scenario needs to be extended to cover anisotropic diffusion. In this paper, we focus on the parallel diffusion with isotropic pitch-angle scattering, which may be an approximation to the transport process driven by weak hydromagnetic waves in a magnetic flux tube, where fluctuations of the wave velocities lead to the turbulent convection. The consequence is the breaking of the statistical isotropy, while the overall shape of the angular power spectrum, $ \overline{C_\ell}\propto\ell ^{-\gamma -1} $ ($ \ell\gg 1 $), remains similar to that in the isotropic diffusion model, where $ \ell $ are degrees of spherical harmonics, and $ \gamma $ is the turbulence spectral index of the convection field. It is then expected that the power-law index of the TeV CR small-scale angular power spectrum can be explained with the Kolmogorov law $ \gamma =5/3 $, irrespective of the background magnetic field to some extent.

\end{abstract}
\section{Introduction}
It has been widely accepted that the propagation of Galactic cosmic rays (CRs) is diffusive. Reference can be made to the point that the observed CR angular distribution is highly isotropic, and the anisotropy is dominated by a dipole moment whose amplitude roughly increases with the particle energy. This energy dependence is consistent with the picture that these charged relativistic particles are randomly scattered by interstellar magnetic irregularities, while the phase of the dipole anisotropy is coincident with our understanding of possible CR sources in the Milky Way galaxy \citep{2016PhRvL.117o1103A,2021PhRvD.104j3013F,2022ApJ...926...41Z,2022MNRAS.511.6218Z,2023ApJ...942...13Q,2024ApJ...962...43L}.

However, the observations have also shown that the CR distribution has structures with angular scales smaller than the dipole one \citep{2011ApJ...740...16A,2014ApJ...796..108A,2016ApJ...826..220A,2018ApJ...865...57A,2019ApJ...871...96A,2024ApJ...961...87C,2025ApJ...981..182A}. Since the standard diffuse anisotropy is purely dipolar, it is intuitive to claim that the small-scale anisotropies are beyond the classical diffusion\footnote{By default, the terms ``diffusion'' and ``convection'' in this paper refer to the processes in position space.} model. Some radical scenarios have ascribed the small-scale hotspots in the CR intensity sky to signals of neutral strangelets \citep{2013PhLB..725..196K} or dark matter annihilation \citep{2013arXiv1307.6537H}. Most existing models are still within the classical electromagnetism, including anisotropic pitch-angle scattering \citep{2010ApJ...721..750M,2017ApJ...835..258G} and nondiffusive transport in concrete magnetic field structures \citep{2012PhRvL.109g1101G,2014Sci...343..988S,2014ApJ...790....5Z,2016ApJ...822..102H,2017ApJ...842...54L,2025ApJ...979..197B}. Remarkably, the observed CR small-scale angular power spectrum may be interpreted as an ensemble average over realizations of magnetic turbulence before significant (pitch-angle) scattering of the particles occurs \citep{2014PhRvL.112b1101A,2015ApJ...815L...2A,2019JCAP...11..048M,2022ApJ...927..110K}.

Nevertheless, \cite{2024ApJ...964L...1Z} recently suggested that the small-scale anisotropies may still be described within the classical diffusion approximation, which corresponds to simple random walks. The key is recognizing that complete relaxation of the particle trajectories can be observed only in the fluid rest frame, while the fluid is generally nonuniform---this is not included in the traditional model. As is known, uniform convection, i.e., the Compton--Getting effect, falls outside the relaxation process, and would give rise to a dipole anisotropy nearly independent of the CR energy \citep{2017PrPNP..94..184A}. Regular nonuniform convection encodes inertial and shear effects, which induce dipole and quadrupole anisotropies proportional to the relaxation time, respectively \citep{2022ApJ...938..106Z}. The authors realized that more complex angular correlations of the particle momenta can be established when the convection becomes irregular, i.e., turbulent. They showed that the observed CR angular power spectra at TeV energies may be direct reflections of interstellar Kolmogorov turbulence.

For the sake of simplicity, the convection scenario proposed by \cite{2024ApJ...964L...1Z} is based on isotropic diffusion, which means that the diffusivities parallel and perpendicular to the background (mean) magnetic field are comparable, and should in principle be valid when magnetic fluctuations on the scale of the unperturbed gyroradius $ r_\text{g} $ are not weaker than the background field. This corresponds to the case where, in the standard Bohm limit, the parallel mean free path $ \lambda _\parallel $ is comparable with $ r_\text{g} $; in the modified Bohm limit arising from nonlinear theories, $ \lambda _\parallel $ may be even smaller \citep{2014ApJ...785...31H}. However, a rule of thumb from studies of the CR energy spectrum and dipole anisotropy is $ \lambda _\parallel\left( \text{TeV}\right) \sim 3\text{ pc}\gg r_\text{g}\left( \text{TeV}\right) \sim 60 $ AU, implying that the isotropic diffusion assumption cannot be exactly satisfied for the TeV CRs. Note that we can still isotropize the diffusion tensor by taking an ensemble average over the background fields, with the isotropized mean free path on the same order as $ \lambda _\parallel $ \citep{2022MNRAS.511.6218Z}. But it is doubtful whether such an ensemble is involved in the existing CR measurements. On the other hand, there are observational indications of an alignment between the TeV dipole anisotropy and the local interstellar magnetic field \citep{2014Sci...343..988S,2019ApJ...871...96A}. Therefore, generalizing the convection model of the higher-order anisotropies to incorporate the anisotropic diffusion is of practical importance. In what follows, we shall focus on the extreme case---purely parallel diffusion---by assuming a strong background magnetic field, corresponding to the CR propagation in a nearly uniform magnetic flux tube.
\section{Convection Anisotropy with Parallel Diffusion}
Let us adopt the classical diffusion picture: in a reference frame where a scattering center is at rest, the scattered particles tend to forget their initial states completely. This relaxation process also establishes the state of convection, i.e., the co-movement of the particles with the scattering center, whose velocity will be denoted by $ \boldsymbol{U} $. Since the scattering centers at different locations can have relative motion, the convection is generally nonuniform, leading to a momentum change $ \Delta\boldsymbol{p}=-p\Delta\boldsymbol{U}/v $ when a particle travels over the (inter-scattering) free path\footnote{This usually differs from the ``mean free path'', whose angular dependence is defined to be integrated out.} $ \Delta\boldsymbol{r} $, where $ \boldsymbol{p} $ and $ v $ are the particle momentum and speed, respectively, and $ \boldsymbol{r} $ is the position vector. The $ \Delta\boldsymbol{r} $ and $ \Delta\boldsymbol{p} $, which are assumed to be small, tend to be washed out in the relaxation process, so the unrelaxed fluctuation over the particle isotropic distribution $ f $ is a time-reversal increment $ \Delta f=-\Delta\boldsymbol{r}\cdot\boldsymbol{\nabla}f-\Delta\boldsymbol{p}\cdot\partial f/\partial\boldsymbol{p} $, which may be interpreted as the steady-state anisotropy under the fluctuation--relaxation equilibrium.

The above arguments form the basic idea of the CR convection--diffusion anisotropy in the fluid rest frame. As mentioned, previous modeling \citep{2022ApJ...938..106Z,2024ApJ...964L...1Z} employed isotropic diffusion, which assumes $ \Delta\boldsymbol{r} $ to be $ \boldsymbol{\lambda}=\tau\boldsymbol{v} $ with $ \tau $ the relaxation time, and $ \boldsymbol{U} $ to be consistent with the background flow velocity $ \boldsymbol{u} $. Now we are more interested in anisotropic diffusion, which breaks the assumption that $ \Delta\boldsymbol{r} $ is aligned with $ \boldsymbol{p} $, and that $ \boldsymbol{U} $ is equivalent to $ \boldsymbol{u} $. It should be emphasized that the position-space anisotropy may not necessarily be followed by the momentum-space one \citep{2017PrPNP..94..184A}; clearly, the CR anisotropy refers to the latter. As an example, it can be proven that, for any spatial distribution under natural boundary conditions, isotropic pitch-angle scattering along with uniform convection allows only a dipole anisotropy in the momentum space \citep{2017ApJ...835..258G}.

For simplicity, let us ignore the particle transport perpendicular to the background magnetic field $ \boldsymbol{B} $ by setting $ B $ to be much stronger than magnetic irregularities on the gyroresonance scale $ 2\pi r_\text{g} $ \citep{1999ApJ...520..204G}. Since $ \Delta\boldsymbol{r} $ is now completely along the background magnetic field lines, $ \boldsymbol{U} $ must also be along the field lines, and the system is cylindrically symmetric. In other words,
\begin{align}
\Delta f=\mu \frac{\boldsymbol{B}}{B}\cdot \left( -\lambda \boldsymbol{\nabla }f+p\frac{\Delta \boldsymbol{u}}{v}\frac{\partial f}{\partial p} \right) ,\label{CDA}
\end{align}
where $ \mu =\cos\vartheta =\left( \boldsymbol{B}/B \right) \cdot \boldsymbol{p}/p $ is the cosine of the pitch angle, and $ \Delta \boldsymbol{u}=\boldsymbol{u}\left( \boldsymbol{r}+\lambda\mu\boldsymbol{B}/B \right) -\boldsymbol{u}\left( \boldsymbol{r} \right) $. Specifically, $ \boldsymbol{u} $ may be characterized as the group velocity of hydromagnetic waves, given that the CR scattering centers are mainly the component waves propagating along the field lines. Note that the first term on the right-hand side of Eq.~\eqref{CDA} represents the standard parallel diffuse anisotropy, and the second term arises from the nonuniform convection, which is the focus of our investigation. For a more careful derivation of Eq.~\eqref{CDA}, one can refer to the method introduced in Appx.~A of \cite{2024ApJ...964L...1Z}, with an additional assumption that the distribution function spatially depends only on coordinates along the field lines.

The CR anisotropy is usually studied with the multipole basis projected on the momentum space. To obtain the multipole expansion of the convection term of $ \Delta f $, the first step is to expand the flow field into
\begin{align}
\Delta \boldsymbol{u}=\sum_{L=0}^{\infty}{\boldsymbol{c}_L\sqrt{\frac{2L+1}{4\pi}}P_L\left( \mu \right)},\label{Du}
\end{align}
where $ P_L\left( \mu\right) $ are the Legendre polynomials. The next step is the dipole--$ 2^L $-pole coupling
\begin{align}
\mu \frac{\boldsymbol{B}}{B}\cdot \Delta \boldsymbol{u}=\sum_{\ell =0}^{\infty}{a_{\ell}\sqrt{\frac{2\ell +1}{4\pi}}P_{\ell}\left( \mu \right)}=\sum_{\ell m}{\mathfrak{A}_{\ell m}Y_{\ell m}\left( -\frac{\boldsymbol{p}}{p} \right)},\label{MC}
\end{align}
where $ Y_{\ell m}\left( -\boldsymbol{p}/p \right) $ are the spherical harmonic functions evaluated at the particle arrival direction ($ -\boldsymbol{p}/p $). Clearly, this is a special case of the Clebsch--Gordan series \citep{2020mqm..book.....S}, from which we get
\begin{align}
a_{\ell}=\frac{\boldsymbol{B}}{B}\cdot \sum_{L=0}^{\infty}{\sqrt{\left( 2L+1 \right) \left( 2\ell +1 \right)}\left( \begin{matrix}
	1&		L&		\ell\\
	0&		0&		0\\
\end{matrix} \right) ^2\boldsymbol{c}_L}=\sqrt{\frac{2\ell +1}{4\pi}}\frac{\mathfrak{A}_{\ell m}}{Y_{\ell m}^{*}\left( -\frac{\boldsymbol{B}}{B} \right)}.\label{al}
\end{align}
According to the selection rules of Wigner's 3j symbols, only terms of $ L=\left| \ell \pm 1 \right| $ survive in the above expression. So $ a_1 $ alone is related to the mean field $ \boldsymbol{c}_0 $, which will be excluded from our analysis for turbulent convection.
\section{Turbulent Convection}
It can be demonstrated that the convection anisotropies with $ \ell >2 $ are significant only if the convection is irregular \citep{2024ApJ...964L...1Z}. We thus introduce the turbulent convection for modeling the small-scale anisotropies, with the implication that the CR convection is determined by the background flow on the CR mean-free-path scale. Since $ \Delta\boldsymbol{u} $ is now a random variable, there is no fixed shape of the induced $ \Delta f $ defined by Eq.~\eqref{CDA}, while its ensemble behavior may be predictable. Hence, to establish the statistics of the distributional anisotropies, we first need to study, in spherical coordinates, the average over realizations of the turbulent flow. This ensemble average will be represented by the overline notation in the following.

Conventionally, the spectrum of turbulence is expressed in terms of the Fourier components of the fluctuation field. By definition, the Fourier kernel of interest is $ e^{i\boldsymbol{k}\cdot\lambda\mu\boldsymbol{B}/B} $, where $ \boldsymbol{k} $ is the wave vector. As our focus is on the turbulent convection anisotropies, we shall assume the pitch-angle scattering process to be isotropic, i.e., the scattering time $ \tau =\lambda /v $ is independent of $ \mu $, otherwise the diffuse anisotropy may also have multipole moments. Using the plane-wave expansion, the spherical harmonic coefficients in Eq.~\eqref{Du} can be expressed as
\begin{align}
\boldsymbol{c}_L=\sqrt{4\pi \left( 2L+1 \right)}i^L\int{j_L\left( \frac{\boldsymbol{B}}{B}\cdot \boldsymbol{k}\lambda \right) \tilde{\boldsymbol{u}}\left( \boldsymbol{k} \right) \frac{d^3k}{\left( 2\pi \right) ^3}},
\end{align}
where $ j_L\left( \xi \right) $ is the spherical Bessel function of the first kind, and $ \tilde{\boldsymbol{u}} $ is the Fourier transform of $ \Delta\boldsymbol{u} $.

To simplify, let us consider homogeneous isotropic turbulence \citep{1982tht..book.....B}, whose general formula of the $ \boldsymbol{k} $-space two-point correlation function is
\begin{align}
\overline{\tilde{\boldsymbol{u}}\left( \boldsymbol{k} \right) \tilde{\boldsymbol{u}}^*\left( \boldsymbol{k}' \right) }=\left( 2\pi \right) ^6\delta ^3\left( \boldsymbol{k}-\boldsymbol{k}' \right) \frac{w\left( k \right)}{4\pi k^2}\left[ \frac{\beta}{2}\left( \mathbf{I}-\frac{\boldsymbol{kk}}{k^2} \right) +\left( 1-\beta \right) \frac{\boldsymbol{kk}}{k^2} \right] ,\label{CF}
\end{align}
where $ \mathbf{I} $ denotes the identity tensor, $ \delta ^3\left( \boldsymbol{k}\right) $ is the (3D) Dirac delta function, $ w\left( k\right) $ is the omnidirectional spectrum, and $ \beta $ is the fraction of the transverse components (with respect to $ \boldsymbol{k} $), with $ \beta =0 $, 2/3 and 1 corresponding to the acoustic, equipartition and incompressible flows, respectively. Then
\begin{align}
\frac{\boldsymbol{B}\boldsymbol{B}}{B^2}:\overline{\boldsymbol{c}_L\boldsymbol{c}_{L'}}=\int_0^{\infty}{\left\{ J_{LL'}^{\left( 0 \right)}\left( k\lambda \right) \frac{\beta\left( k \right)}{2}+J_{LL'}^{\left( 2 \right)}\left( k\lambda \right) \left[ 1-\frac{3}{2}\beta\left( k \right) \right] \right\} w\left( k \right) dk},\label{cLcL'}
\end{align}
where the ``$ : $'' symbol between two dyadics represents the double-dot product, and
\begin{align}
J_{LL'}^{\left( n \right)}\left( \xi \right) =2\pi \sqrt{\left( 2L+1 \right) \left( 2L'+1 \right)}i^{L-L'}\int_{-1}^1{j_L\left( \xi x \right) j_{L'}\left( \xi x \right) x^ndx},\label{ILL'}
\end{align}
which is compatible with real $ \overline{\boldsymbol{c}_L\boldsymbol{c}_{L'}} $ because the parity $ j_L\left( -\xi \right) =\left( -1 \right) ^Lj_L\left( \xi \right) $ leads to vanishing integral results for $ L+L'+n $ being odd. For the physical system of interest, $ L $, $ L' $ and $ n $ are all nonnegative integers.

In order to find the analytic solution, we note that the Bessel function can also be expressed in terms of the generalized hypergeometric function $ {}_0F_1 $, and the product of two $ {}_0F_1 $ functions is equivalent to one $ {}_2F_3 $. Then, the nonzero branch of the integral in Eq.~\eqref{ILL'} can be seen as a special case of the Euler-type transform of $ {}_2F_3 $. Accounting for the zero branch, one has
\begin{align}
\int_{-1}^1{j_L\left( \xi x \right) j_{L'}\left( \xi x \right) x ^ndx}=\frac{2^{-L-L'-2}\pi \left[ 1+\left( -1 \right) ^{L+L'+n} \right] \xi ^{L+L'}}{\left( L+L'+n+1 \right) \Gamma \left( L+\frac{3}{2} \right) \Gamma \left( L'+\frac{3}{2} \right)}{}_3F_4\left(\substack{\frac{L+L'}{2}+1,\frac{L+L'+3}{2},\frac{L+L'+n+1}{2}\\L+\frac{3}{2},L'+\frac{3}{2},L+L'+2,\frac{L+L'+n+3}{2}};-\xi ^2 \right) ,
\end{align}
where $ \Gamma\left( \xi\right) $ is the gamma function. For details, see \cite{Slater2008}.
\section{Statistics of the Anisotropies}
\begin{figure}
	\centering
	\includegraphics[width=0.5\textwidth]{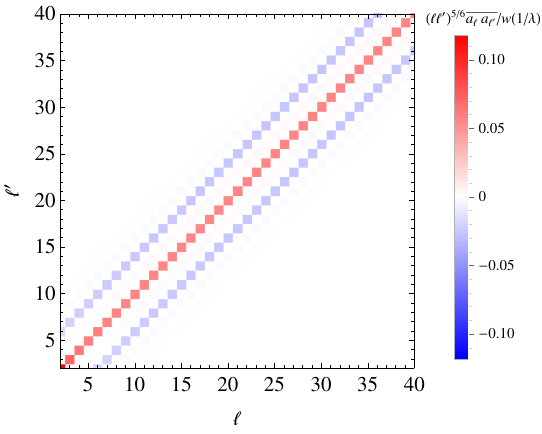}
	\caption{Angular correlation matrix of the turbulent convection anisotropies for $ \gamma =5/3 $ and $ \beta =2/3 $.}\label{f1}
\end{figure}
With Eq.~\eqref{al} and the definition of the 3j symbols, the explicit result for the two-mode correlation of the particle angular distribution can be given by
\begin{align}
\overline{a_{\ell}a_{\ell '}}=\frac{1}{\sqrt{\left( 2\ell +1 \right) \left( 2\ell '+1 \right)}}\frac{\boldsymbol{B}\boldsymbol{B}}{B^2}:&\left[ \frac{\ell \ell '\overline{\boldsymbol{c}_{\ell -1}\boldsymbol{c}_{\ell '-1}}}{\sqrt{\left( 2\ell -1 \right) \left( 2\ell '-1 \right)}}+\frac{\ell \left( \ell '+1 \right) \overline{\boldsymbol{c}_{\ell -1}\boldsymbol{c}_{\ell '+1}}}{\sqrt{\left( 2\ell -1 \right) \left( 2\ell '+3 \right)}} \right. \notag\\
&\left. +\frac{\left( \ell +1 \right) \ell '\overline{\boldsymbol{c}_{\ell +1}\boldsymbol{c}_{\ell '-1}}}{\sqrt{\left( 2\ell +3 \right) \left( 2\ell '-1 \right)}}+\frac{\left( \ell +1 \right) \left( \ell '+1 \right) \overline{\boldsymbol{c}_{\ell +1}\boldsymbol{c}_{\ell '+1}}}{\sqrt{\left( 2\ell +3 \right) \left( 2\ell '+3 \right)}} \right] ,
\end{align}
which is visualized by Fig.~\ref{f1}. As seen, this correlation matrix differs from the standard diagonal form in the isotropic diffusion model\footnote{For the isotropic diffusion model with the equipartition assumption ($ \beta =2/3 $), \cite{2024ApJ...964L...1Z} pointed out that the distributional angular correlation is determined by $ \overline{a_{\ell m}a_{\ell 'm'}^{*}}=\left( A_{\ell}\delta _{\ell \ell '}+\mathcal{A}_{\ell m}\delta _{\ell ,\ell '+2}+\mathcal{A}_{\ell 'm}\delta _{\ell ',\ell +2} \right) \delta _{mm'} $, which can be inferred from the selection rules of the 3j symbols. In fact, the orthogonality of the symbols also requires $ \mathcal{A}_{\ell m}=0 $.}, indicating that the statistical isotropy can be broken by the background magnetic field. Nevertheless, the angular power spectrum is defined to be not associated with the off-diagonal elements \citep{2017PrPNP..94..184A}; for the convection term in Eq.~\eqref{CDA}, the ensemble-averaged spectrum is
\begin{align}
\overline{C_\ell}=\left( \frac{1}{v}\frac{\partial \ln f}{\partial \ln p} \right) ^2\frac{\overline{a_{\ell}^2}}{2\ell +1}.\label{Cl}
\end{align}

Note that a simple approach to producing an isotropic random system from the anisotropic diffusion model is to introduce an ensemble of the background fields. To show this, it is first necessary to decouple the coordinate system of observation from $ \boldsymbol{B}/B $, as has been done (with the spherical harmonic addition theorem) in the last step of Eq.~\eqref{MC}. Then, due to the orthogonality of spherical harmonics, the average over $ \boldsymbol{B}/B $ would lead to
\begin{align}
\left< \mathfrak{A}_{\ell m}\mathfrak{A}_{\ell 'm'}^{*} \right> _{\frac{\boldsymbol{B}}{B}}=\frac{a_{\ell}^{2}}{2\ell +1}\delta _{\ell \ell '}\delta _{mm'},\label{IRF}
\end{align}
with $ \delta _{\ell\ell '} $ the Kronecker delta. This formally coincides with the covariance of the statistically isotropic fluctuations.

Further assumptions about the turbulence spectrum are necessary for deriving more specific forms of the angular correlation. It can be shown that $ J_{LL'}^{\left( n \right)}\left( \xi \right) \propto\xi ^{L+L'} $ as $ \xi\rightarrow 0 $, and the slowest decay follows $ J_{LL'}^{\left( n \right)}\left( \xi \right)\propto\mathcal{C} _{LL'}^{\left( n \right)}\xi ^{-n-1}+\xi ^{-2} $ ($ \mathcal{C} _{LL'}^{\left( n \right)} $ is $ \xi $-independent) as $ \xi\rightarrow\infty $. Thus, for a moderately flat $ w\left( k\right) $, which is the feature of usual cascade processes, the limiting behaviors of the integrand at small and large $ k $ in Eq.~\eqref{ILL'} are not important for the integral. As we are interested in CRs at 10 TeV, while the Galactic turbulence injection scale may be larger than $ \lambda\left( 10\text{ TeV}\right) \sim 10 $ pc \citep{2010ApJ...710..853C}, we can consider only the inertial range for simplicity, i.e, $ w\left( k\right) \propto k^{-\gamma} $ with $ k $-independent $ \gamma\sim 1 $. The convergence condition is then $ -\min \left( n,1 \right) <\gamma <L+L'+1 $, under which, according to the Merlin transform of $ {}_3F_4 $ \citep{Prudnikov1989}, the exact solution is
\begin{align}
\int_0^{\infty}{J_{LL'}^{\left( n \right)}\left( k\lambda \right) w\left( k \right) dk}=\frac{\pi ^2\sqrt{\left( 2L+1 \right) \left( 2L'+1 \right)}i^{L-L'}\left[ 1+\left( -1 \right) ^{L+L'+n} \right] w\left( \frac{1}{\lambda} \right) \Gamma \left( \gamma +1 \right) \Gamma \left( \frac{L+L'-\gamma +1}{2} \right)}{2^{\gamma +1}\left( \gamma +n \right) \lambda \Gamma \left( \frac{L-L'+\gamma}{2}+1 \right) \Gamma \left( \frac{L'-L+\gamma}{2}+1 \right) \Gamma \left( \frac{L+L'+\gamma +3}{2} \right)},
\end{align}
which can be used to simplify Eq.~\eqref{cLcL'} if $ \beta $ is independent of $ k $. This also shows that $ n $ with a given parity is not coupled with $ L $ and $ L' $, so $ \beta $ affects only the normalization factor of the multipole spectrum.

In the large-$ L $ limit, one has
\begin{align}
\frac{\boldsymbol{B}\boldsymbol{B}}{B^2}:\overline{\boldsymbol{c}_L\boldsymbol{c}_L}=\left( \frac{\beta}{\gamma}+\frac{2-3\beta}{\gamma +2} \right) \frac{\pi ^{\frac{3}{2}}\Gamma \left( \frac{\gamma +1}{2} \right)}{\lambda \Gamma \left( \frac{\gamma}{2}+1 \right)}w\left( \frac{L}{\lambda} \right) =-\left( 1+\frac{2}{\gamma} \right) \frac{\boldsymbol{B}\boldsymbol{B}}{B^2}:\overline{\boldsymbol{c}_L\boldsymbol{c}_{L+2}}.
\end{align}
Subsequently, Eq.~\eqref{Cl} yields $ \overline{C_\ell}\propto p^{\left( \gamma -1 \right) \left( 2-\gamma _{\text{g}} \right)}\ell ^{-\gamma -1} $, which is identical to the isotropic diffusion model for $ \ell\gg 1 $ \citep{2024ApJ...964L...1Z}. Note that the energy dependence arises partially from the empirical formula $ \lambda\propto p^{2-\gamma _{\text{g}}} $; for the quasilinear scattering theory \citep{1966ApJ...146..480J,2008ApJ...685L.165T}, $ \gamma _{\text{g}} $ can be interpreted as the spectral index of slab turbulence at the gyroresonance wavenumber $ 1/r_{\text{g}} $. As $ r_{\text{g}}\ll\lambda $, the turbulence inducing the scattering can have very different symmetry from that defining the convection field.

\begin{figure}
	\centering
	\includegraphics[width=0.5\textwidth]{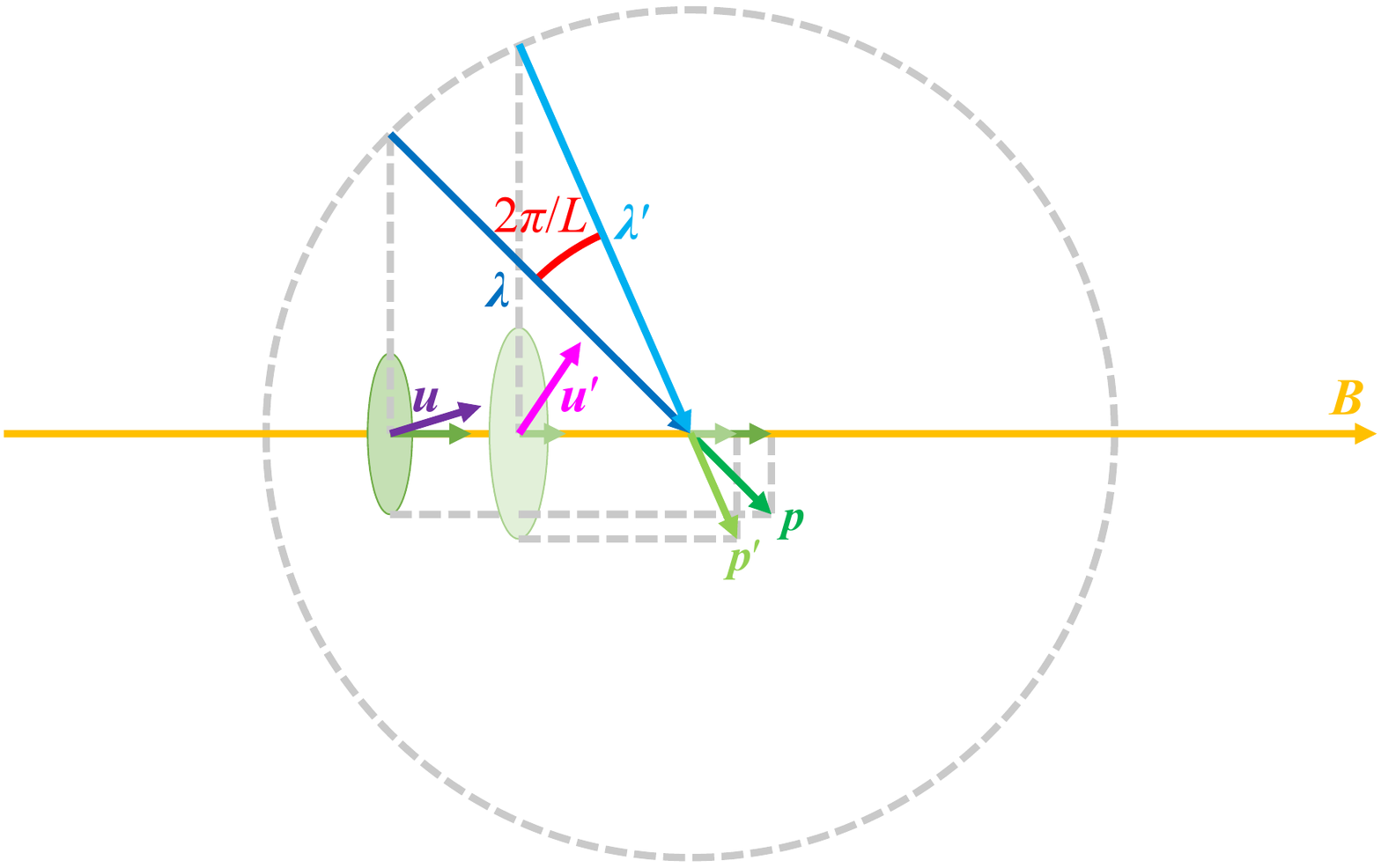}
	\caption{Schematic view of the field-aligned nonuniform convection in the isotropic pitch-angle scattering regime.}\label{f2}
\end{figure}

Given the orthogonality of $ j_L\left( \xi \right) $, the large-$ \ell $ behavior of $ \overline{C_\ell} $ may also be understood via the approximation
\begin{align}
j_L\left( \xi \right) j_{L'}\left( \xi \right) \sim \frac{\pi}{2\left( 2L+1 \right)}\delta _{LL'}\delta \left( \left| \xi \right|-L \right) ,\label{jLjL'}
\end{align}
from which the integral in Eq.~\eqref{ILL'} reduces to $ \pi \left[ 1+\left( -1 \right) ^n \right] L^n\delta _{LL'}\Theta \left( \left| \xi \right|-L \right) /\left[ 2\left( 2L+1 \right) \left| \xi \right|\xi ^n \right] $, where $ \Theta\left( \xi\right) $ is the unit step function, implying that the integral in Eq.~\eqref{cLcL'} is dominated by the integrand at $ k\sim L/\lambda $ for $ \gamma\sim 1 $. The underlying picture is that the difference in the convection properties of two particles is determined by the separation of the particles' initial positions, as shown by Fig.~\ref{f2}. Note that the angular scale of the $ 2^L $-pole moment is $ 2\pi /L $, corresponding to a length scale $ 2\pi\lambda /L $ on a sphere of radius $ \lambda $. Although $ \boldsymbol{\lambda} $ is now not the true free path for the particles in spiral motion around the background magnetic field lines, the $ \boldsymbol{B} $-field projection does not change the order of magnitude of the length scale (despite an average reduction by a factor of $ \pi /2 $). Therefore, the characteristic scale of the convection property variation determining the $ 2^\ell $-pole anisotropy is always on the order of $ 2\pi\lambda /\ell $. We can then expect that the $ \ell $ and energy dependences of $ \overline{C_\ell} $ in the parallel diffusion model are similar to those in the isotropic diffusion model.
\section{Comparison with Observations}
\begin{figure}
	\centering
	\includegraphics[width=\textwidth]{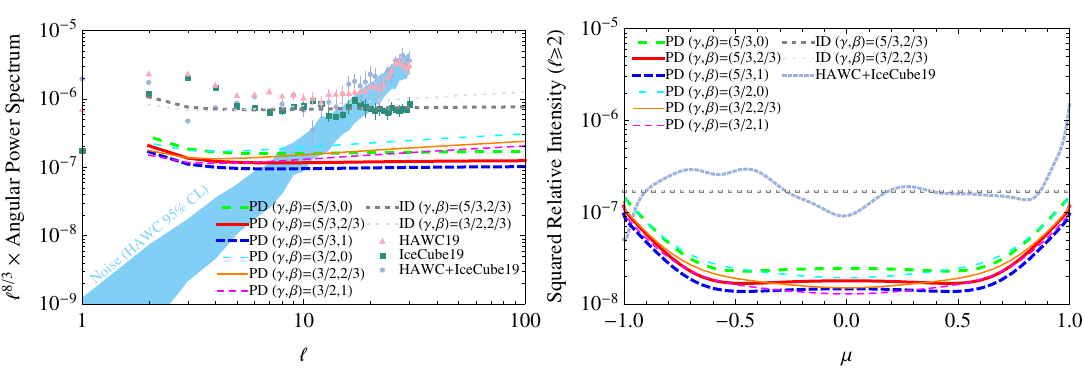}
	\caption{Comparison of the isotropic diffusion \citep[ID;][]{2024ApJ...964L...1Z} and parallel diffusion (PD) models of the turbulent convection anisotropies, with $ \sigma\left( 10\text{ TeV}\right) \approx 23 $ km/s and $ \partial\ln f/\partial\ln p=-4.7 $. Left: The angular power spectra. Note that the thick dashed line in dark gray represents the fit to the CR observational data at 10 TeV \citep{2019ApJ...871...96A} in the small-scale range. Right: The squared angular distributions for $ \ell\geqslant 2 $, normalized by the isotropic background.}\label{f3}
\end{figure}

The above large-$ \ell $ behavior of $ \overline{C_\ell} $, along with the Kolmogorov law\footnote{As the existing data involve only $ \ell\lesssim 30 $, the Kraichnan law $ \gamma =3/2 $ may also work, but with a slightly worse fitting performance.} $ \gamma =5/3 $, has been demonstrated in previous work \citep{2024ApJ...964L...1Z} to be consistent with the CR small-scale angular power spectrum observed by HAWC and IceCube at 10 TeV up to $ \ell =30 $. However, as shown in the left-hand panel of Fig.~\ref{f3}, for a given isotropic configuration of the background turbulence on the scale of $ 2\pi\lambda /\ell $, the parallel diffusion model predicts a $ \overline{C_\ell} $ several times lower than that predicted by the isotropic diffusion model. This is because we have assumed the particle convection in the former model to be field-aligned so that the cross-field background modes with $ \tilde{\boldsymbol{u}}\cdot\boldsymbol{B}=0 $, which contribute to $ w\left( k\right) $, have no contribution to $ \overline{C_\ell} $; furthermore, there is a $ \boldsymbol{B} $-field projection of the characteristic scale. Note that the isotropic diffusion model would require $ \sigma\left( 10\text{ TeV}\right) \sim 20 $ km/s (for $ \lambda\left( 10\text{ TeV}\right) \sim 10 $ pc) to account for the observations, where $ \sigma =\sqrt{2\pi w\left( 2\pi /\lambda \right) /\lambda} $ is the velocity dispersion of the background flow (plasma waves) on the scale of $ \lambda $. To fit the observational data with the parallel diffusion model, we need to set $ \sigma\left( 10\text{ TeV}\right) \sim 50 $ km/s. There would be more flexible choices of this value if we took an anisotropic form for Eq.~\eqref{CF}.

By definition, $ \overline{C_\ell} $ does not reflect the symmetry of the system. We have previously mentioned that the turbulent convection anisotropies in the anisotropic diffusion model should be statistically anisotropic. To further visualize, in the right-hand panel of Fig.~\ref{f3} we plot the mean-square angular distribution corresponding to the total turbulent convection anisotropy, i.e.,
\begin{align}
\frac{\overline{\left(\Delta f\right)^2_\text{cov}}}{f^2}=\left( \frac{1}{v} \frac{\partial\ln f}{\partial\ln p} \right)^2\sum _{\ell, \ell' =2}^{\ell_{\max}}{\sum _{mm'}{\overline{a_{\ell m}a_{\ell 'm'}^*} Y_{\ell m}\left( \frac{\boldsymbol{p}}{p} \right) Y_{\ell 'm'}^*\left( \frac{\boldsymbol{p}}{p} \right)}}.\label{TAP}
\end{align}
For the isotropic diffusion model, the summation result is independent of $ \boldsymbol{p}/p $ because $ \overline{a_{\ell m}a_{\ell 'm'}^*}=A_\ell\delta _{\ell \ell '}\delta _{mm'} $. For the parallel diffusion model, one has $ \overline{a_{\ell m}a_{\ell 'm'}^*}=\overline{a_\ell a_{\ell '}}\delta _{m0}\delta _{m'0} $ in the magnetic coordinate system, whose polar axis is aligned with $ \boldsymbol{B} $. We take $ \ell _{\max}=50 $ to compute the models. It shows that $ \overline{\left(\Delta f\right)^2_\text{cov}} $ tends to increase as $ \boldsymbol{p} $ approaches the $ \pm\boldsymbol{B} $ directions.

On the other hand, to deal with the CR relative intensity $ \mathcal{I} =\sum_{\ell m}{\mathfrak{a}_{\ell m}Y_{\ell m}\left( \alpha ,\pi /2-\delta \right)} $ observed to be not fully cylindrically symmetric, where $ \alpha $ and $ \delta $ are the right ascension and declination of the particle arrival direction, respectively, we first rotate the observational coordinate system from the equatorial to the magnetic one, and then average $ \mathcal{I}^2 $ over the azimuth angle $ \varphi $ around $ \boldsymbol{B} $ to extract the dependence on the pitch angle $ \vartheta $. The quantity of interest is then
\begin{align}
&\left< \mathcal{I} _{\ell \geqslant 2}^{2} \right> _{\varphi}\notag \\
&=\sum_{\ell ,\ell '=2}^{\ell _{\max}^{\mathrm{obs}}}{\left( -1 \right) ^{\ell +\ell '}\sum_{mm'm''}{\mathfrak{a}_{\ell m}\mathfrak{a}_{\ell 'm'}^{*}D_{mm''}^{\left( \ell \right) *}\left( \alpha _{\boldsymbol{B}},\frac{\pi}{2}-\delta _{\boldsymbol{B}},0 \right) D_{m'm''}^{\left( \ell \right)}\left( \alpha _{\boldsymbol{B}},\frac{\pi}{2}-\delta _{\boldsymbol{B}},0 \right) Y_{\ell m''}\left( \vartheta ,0 \right) Y_{\ell 'm''}\left( \vartheta ,0 \right)}},\label{TAPO}
\end{align}
with $ D_{mm'}^{\left( \ell\right)}\left( \phi ,\theta ,\psi\right) $ the Wigner D function \citep{2020mqm..book.....S}. This is also plotted in the right-hand panel of Fig.~\ref{f3}, where $ \mathfrak{a}_{\ell m} $ are taken from (Tab.~3 of) \cite{2019ApJ...871...96A}, which gives $ \ell_{\max}^\text{obs}=14 $; for the equatorial coordinates of the local interstellar magnetic field direction, we use the ``boundary fit'' result $ \left( \alpha _{\boldsymbol{B}},\delta _{\boldsymbol{B}}\right) =\left( 229.2,11.4\right) ^\circ $. As seen, the profile of the observed relative intensity for $ \ell\geqslant 2 $ indeed bears a resemblance to our theoretical predictions, especially for the beam-like structure around $ \vartheta =0 $. Nevertheless, no model can perfectly reproduce the observed complicated morphology, seemingly implying, e.g., the necessity of asymmetric pitch-angle scattering \citep{2015PhPl...22i1504M,2015ApJ...811....8W}.

This discrepancy might not be such a serious issue either, in the sense that the observed intensity sky is more likely to be a particular realization of a fluctuating map, instead of the ensemble-averaged one. In other words, Eq.~\eqref{TAP} is the expectation value of Eq.~\eqref{TAPO}; they are not necessarily the same. Meanwhile, one can generally expect the ensemble-averaged angular power spectrum $ \overline{C_\ell} $ to be comparable to the observed, seemingly non-ensemble-averaged spectrum $ C_{\ell}=\sum_{m=-\ell}^{\ell}{\left| \mathfrak{a}_{\ell m} \right|^2}/\left( 2\ell +1 \right) $ in the small-scale range, because the latter, by definition, has in fact implicitly included the ensemble nature---an average over $ 2\ell +1 $ degrees of freedom of the spherical harmonic components $ \mathfrak{a}_{\ell m} $, which can also fluctuate (with a covariance, e.g., in the form of Eq.~\ref{IRF}) according to the ergodic hypothesis.

However, the parallel diffusion model has effectively only one degree of freedom, i.e., $ m\equiv 0 $ (in the magnetic coordinate system), leading to a theoretical $ C_\ell $ virtually impossible to approach the smoothed form of $ \overline{C_\ell} $, even in the small-scale range. This problem could be addressed if we took into account the perpendicular transport, e.g., perpendicular diffusion and transverse density-gradient drift \citep{1975Ap&SS..32...77F,2020SSRv..216...23S}, as the cross-field particle displacements can break the cylindrical symmetry of the turbulent convection, and then restore the missing $ 2\ell $ distinct values of $ m $. Intuitively, for two incoming equal-energy particles with their momenta separated by an angle $ 2\pi /\ell $ (see Fig.~\ref{f2}), the initial position separation should have comparable contributions from the parallel and perpendicular directions when $ \ell $ is sufficiently large. For smaller $ \ell $, the parallel component dominates if the parallel transport is much faster than the perpendicular one. Consequently, in both cases, the characteristic scale of the turbulence power determining the strength of the $ 2^\ell $-pole anisotropy is on the order of $ 2\pi\lambda /\ell $ (see the arguments right below Eq.~\ref{jLjL'}). From this perspective, the overall behavior of $ \overline{C_\ell} $ with respect to $ \ell $ might not be changed significantly by the incorporation of the perpendicular transport effects.
\section{Summary}
In this paper, we investigate the effect of a nonzero background magnetic field on the turbulent convection scenario of CR small-scale anisotropies, with the focus on parallel diffusion as interstellar magnetic irregularities on the CR gyroradius scales at TeV energies are expected to be small. It is found that, even for isotropic pitch-angle scattering, which would not produce higher-order anisotropies beyond the dipole one if nonuniform convection were not included, the particle angular distribution can exhibit significant multipole anisotropies because of the field-aligned turbulent convection. For a power-law turbulent convection spectrum $ \propto k^{-\gamma} $, the ensemble-averaged angular power spectrum of the small-scale anisotropies in the parallel diffusion model satisfies $ \overline{C_\ell}\propto\ell ^{-\gamma -1} $, which gives a spectral index identical to that in the isotropic diffusion model for $ \ell\gg 1 $. It is reasonable to speculate that a similar $ \ell $ dependence of $ \overline{C_\ell} $ holds for more general anisotropic diffusion. Then, independent of the structure of the diffusion tensor, the observed CR angular power spectrum at 10 TeV implies that interstellar Alfv\'enic convection follows the Kolmogorov law $ \gamma =5/3 $. The data fit with the parallel diffusion model gives a velocity dispersion of about 50 km/s on the scale of 10 pc, provided that the interstellar turbulence on scales around 1--20 pc (corresponding to $ \ell =30 $--$ 2 $) is homogeneous and isotropic.

Although the background magnetic field appears to have a minor impact on the scaling exponent of $ \overline{C_\ell} $ over a broad range of $ \ell $, it controls the symmetry of the detailed momentum distribution. For instance, the parallel diffusion, which falls within the strong-field limit, allows only the cylindrical symmetry. Furthermore, the corresponding turbulent convection anisotropies are statistically anisotropic. Our computation suggests that these anisotropies tend to fluctuate more violently in the field-aligned than in the cross-field directions, showing potential consistency with the observed CR pitch-angle distribution.
\section*{Acknowledgment}
This work is supported by grants from the National Natural Science Foundation of China (Nos.~12303055, 12375103, U1931204), and the Fundamental Research Funds for the Central Universities (2682025CX030).
\bibliographystyle{aasjournal}
\bibliography{ref}
\end{document}